\documentclass[english,keywords,amsmath,amssymb,showpacs]{revtex4}
\usepackage[T1]{fontenc}
\usepackage[latin1]{inputenc}
\usepackage{babel}%
\usepackage{graphicx}
\usepackage{color}
\usepackage{bm}
\usepackage{longtable}
\usepackage{amsmath}
\usepackage{amsfonts}
\usepackage{amssymb}

\begin{document}
\title{Sub-Planck structure in a mixed state}
\author{Asmita Kumari$^1$}
\author{A. K. Pan$^{1}$ \footnote{akp@nitp.ac.in}}
\author{P. K. Panigrahi$^{2}$}
\affiliation{$^{1}$ National Institute Technology Patna, Ashok Rajhpath, Patna, Bihar 800005, India}
\affiliation{$^{2}$Indian Institute of Science Education and Research Kolkata, Mohanpur, Nadia 741246, India}
\begin{abstract}
The persistence of sub-Planck structure in phase space with loss of coherence is demonstrated in a mixed state, which comprises two terms in the density matrix. Its utility in carrying out Heisenberg-limited measurement and quantum parameter estimation have been shown. It is also shown that the mixed state performs equally well as the compass state for carrying out precision measurements. The advantage of using mixed state relies on the fact that such a state can be easier to prepare and may appear from pure states after partial loss of coherence.  We explicate the effect of environment on these sub-Planck structures in the mixed state and estimates the time scale of complete decoherence.  
\end{abstract}
\maketitle
\section{Introduction}

Measurement in quantum mechanics (QM) is in marked contrast to the classical paradigm due to the invasive character of the measurement process in quantum mechanics. The precession of the measurement in quantum world is not only dependent on technology, but also on the inherent fundamental constraints imposed by the structure of the theory itself. Such a constraint was first put forward by Heisenberg by stating that the product of uncertainties for the simultaneous measurements of position and momentum has a lower bound of $\hbar/2$. Heisenberg limited measurement has  recently received upsurge of interest to determine how accurately a small parameter can be estimated with the aid of QM principles. It is a primary task of quantum metrology to find states suitable for improved precision measurement. In a seminal paper, Zurek \cite{zurek}  first demonstrated that for a compass state (a superposition of four suitable minimum-uncertainty Gaussian states) a subtle interference effect in the phase space occurs, leading to a curious structure, which he christened as sub-Planck  structure. The Wigner function of the compass state exhibits oscillations due to interference on a scale of action that can be much less than $\hbar$. In contrast to the commonly believed notion that the phase-space structures smaller than $\hbar$ have no observable consequence, Zurek showed that such a structure enhances the sensitivity of a quantum state to an external perturbation. 

Later, it is found that similar effect can also occur in cat states \cite{toscano,prx1}. A classical wave optics analogue has been tested experimentally \cite{pryx}  in the  time-frequency domain of an electromagnetic field wave packet. Sub-Planck structure and its implications for different physical systems have also been investigated \cite{roy}. A number of proposals have been advanced for generating single particle cat and generalized states, showing the above feature    \cite{toscano,gsa}. A connection between the sub-Planck structure and the quantum weak value has also been proposed \cite{pp}. Recently, the superposition of two or more coherent states has been experimentally produced to demonstrate the quantum state collapse and revival due to the single-photon Kerr effect \cite{kir}. 

Evidently, the extreme sensitivity of these states to small changes on co-ordinate and momenta can make them unstable for quantum parameter estimations. In fact, Zurek \cite{zurek} has shown that the environment induced phase shifts to a quantum state at the sub-Planck action scale, can cause orthogonality between perturbed and unperturbed compass states to drive decoherence.  Hence, it is of deep interest to investigate the persistence of this sub-Planck structure in mixed states affected by loss of coherence. In this paper, we show that the sub-Planck structure manifests in a  minimalist mixed state, containing only \emph{two terms} in the density matrix, as compared to six terms in the pure compass state. This clearly brings out the characteristic interference required for parameter estimation. The compass state is more fragile to decoherence than the mixed state used in this paper. We carefully study how the mixed state performs in estimating small parameters and find that it provides similar sensitivity of the compass state. We outline the procedure to produce the proposed state and explicate how decoherence further affects the proposed mixed state leading to complete decoherence. The decoherence time is calculated, which is same as the compass state. It is worth emphasizing that the advantage of using mixed state lies in it being easier to prepare than the compass state.

The paper is organized as follows. In Section II, we explicate how a mixed state with only two terms in the density matrix produces the sub-Planck structure. The utility of such a state for precision measurement is discussed in Section III in comparison to the compass state. Section IV deals with the effect of decoherence on the  mixed state, where we compute the time scale of complete decoherence. We conclude in Section V after a brief summary of our findings.  

\section{Sub-planck structure of mixed state}
Sub-Planck structure was first demonstrated by Zurek \cite{zurek} by using the superposition of four coherent states. Two of these state localized in the co-ordinate space and the other two are located in momentum space. Subtle effects of interference amongst various terms play a key role in generating sub-Planck structure. It is natural to enquire about the persistence of these structures in the case of more naturally occurring mixed states. In order to show how mixed state produces the sub-Planck structure, we first introduce the relevant density matrix that is required for our purpose. The density matrix $\rho$ can be written in position representation as, 
\begin{equation}
\rho=|\psi\rangle\langle\psi|=\int_{-\infty}^{\infty}dx^{\prime}\int_{-\infty}^{\infty}dx^{\prime\prime} 
\psi^{*}(x^{\prime\prime})\psi(x^{\prime})|x^{\prime}\rangle\langle x^{\prime\prime}|
\end{equation}

If the initial state is taken to be an incoherent mixture of two cat states, viz., $|\psi_{c_1}\rangle=\int_{-\infty}^{\infty} \psi_{c_1}(x)|x\rangle dx$ and $|\psi_{c_2}\rangle=\int_{-\infty}^{\infty}\psi_{c_2}(x) |x\rangle dx$, the density matrix of the incoherent mixture can be written as $\rho=1/2\left( |\psi_{c_1}\rangle\langle\psi_{c1}|+|\psi_{c_2}\rangle\langle\psi_{c_2}|\right)$. In position representation, 
 \begin{eqnarray}
\label{wf1}
\rho&=&\int_{-\infty}^{\infty}dx^{\prime}\int_{-\infty}^{\infty}dx^{\prime\prime} 
\psi_{c_1}^{*}(x^{\prime\prime})\psi_{c_1}(x^{\prime})|x^{\prime}\rangle\langle x^{\prime\prime}|+\int_{-\infty}^{\infty}dx^{\prime}\int_{-\infty}^{\infty}dx^{\prime\prime} 
\psi_{c_2}^{*}(x^{\prime\prime})\psi_{c_2}(x^{\prime})|x^{\prime}\rangle\langle x^{\prime\prime}|,
 \end{eqnarray}
where elements of the density matrix are $\langle x^{\prime\prime}|\rho|x^{\prime}\rangle=\psi^{*}(x^{\prime\prime})\psi(x^{\prime})$ 
with $Tr(\rho)=\int_{-\infty}^{\infty} \langle x|\rho|x\rangle dx=1$. 
 Since we are dealing with a continuous variable system, it is convenient to use the Wigner representation of the density matrix:
\begin{equation}
W(x,p)=\int_{-\infty}^{\infty} \langle x- \frac{y}{2}|\rho|x+\frac{y}{2}\rangle e^{i p y/\hbar} dy
\end{equation} 
\begin{figure}[h]
{\rotatebox{0}{\resizebox{10.0cm}{7.0cm}{\includegraphics{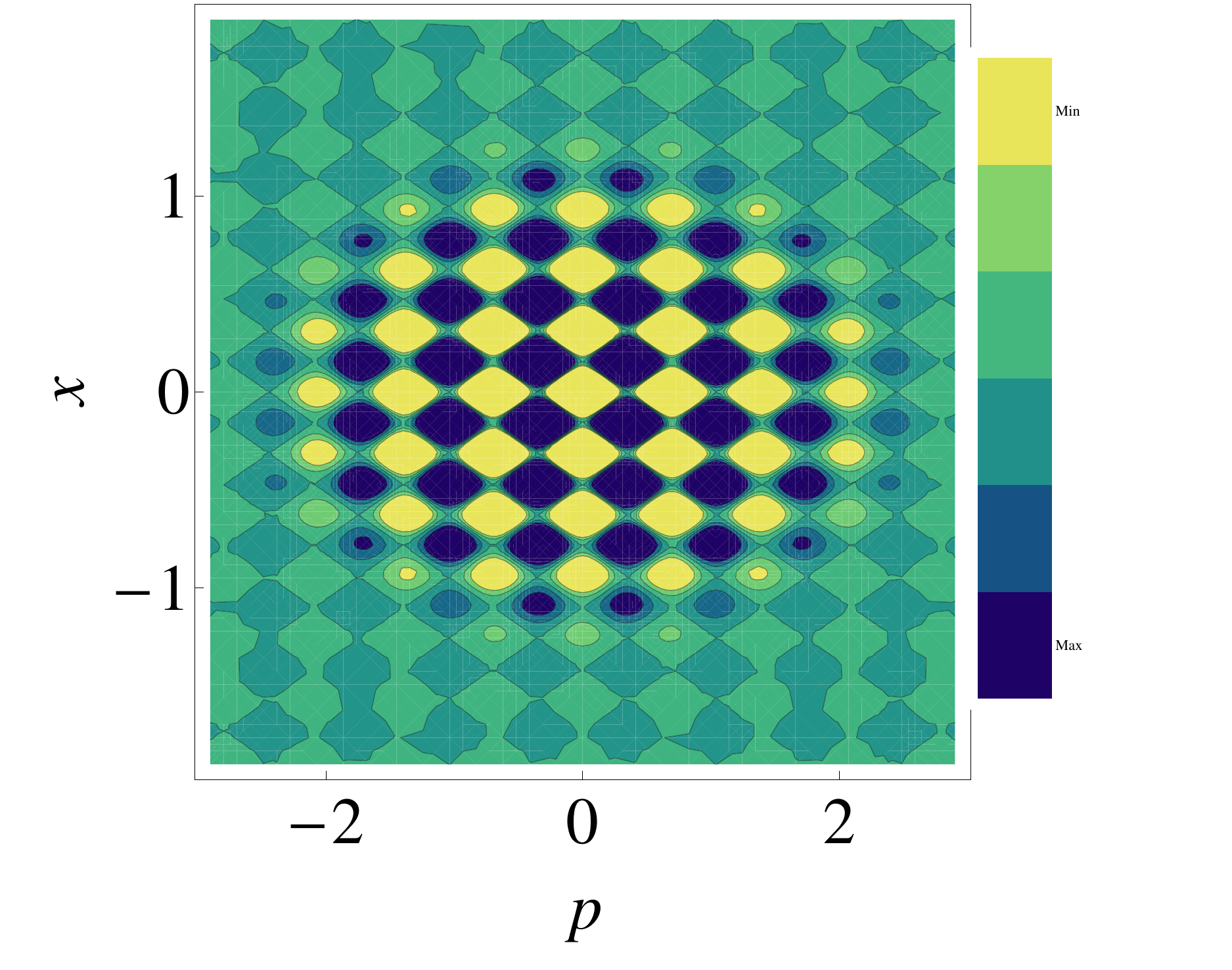}}}}
\caption{(colour online)The Wigner function $W_{\rho}$  given by Eq.(\ref{wrho}) is plotted exhibiting a chess-board type structure. The values of the relevant parameters  are $\sigma=0.5 cm$, $ x_{0}=4.5 cm$ and $p_{0}=10 gm.cm/s$.}
\end{figure}

For the present purpose, we consider two cat states, both of which are superposition of two Gaussian wavefunctions;
\begin{eqnarray}
\label{psic1}
\psi_{c_1}(x)= N_{1}\left[(2\pi\sigma^2)^{-1/4} e^{-\frac{(x - x_{0})^2}{4 \sigma ^2}}+(2\pi\sigma^2)^{-1/4} e^{-\frac{(x + x_{0})^2}{4 \sigma ^2}}\right]
\end{eqnarray}
and
\begin{eqnarray}
\label{psic2}
\psi_{c_2}(x)= N_{2}\left[(2\pi\sigma^2)^{-1/4}e^{-\frac{x^2}{4 \sigma ^2} + \frac{i p_{0} x}{\hbar}}+(2\pi\sigma^2)^{-1/4} e^{-\frac{x^2}{4 \sigma ^2}  -\frac{i p_{0} x}{\hbar}}\right]
\end{eqnarray}
where $\sigma$, $ x_{0}$, and $p_{0}$ are the initial width, initial peak position and initial average momentum of the single Gaussian respectively, and $N_{1}=1/\sqrt{2(e^{-\frac{{x_0}^2}{2 \sigma ^2}}+1) }$ and $N_{2}=1/\sqrt{2(e^{-\frac{2 {p_0}^2 \sigma ^2}{\hbar ^2}}+1)}$ are the normalization constants. \\
 
For a general mixed state, $\rho=\sum_{i}P_{i}\rho_{i}$, where $\rho_i$ is the the $i$-th density matrix having probability $P_i$, the Wigner function follows the distributive property, $W = \sum_{i} P_{i} W_{i}$, using which we calculate the Wigner function for the density matrix $\rho$, given by Eq.(\ref{wf1}),  so that, $W_{\rho} =1/2\left(W_{c_1}+W_{c_2}\right)$, where $W_{c_1}$ and $W_{c_2}$ are given by 
\begin{equation}
\label{wc1}
W_{c_{1}}=N_{1}^{2}\frac{e^{-\frac{2 p^2 \sigma ^2}{\hbar ^2}-\frac{(x+{x_{0}})^2}{2 \sigma ^2}} \left[e^{\frac{{x_{0}} (2 x+{x_{0}})}{2 \sigma ^2}} \cos \left(\frac{2 p {x_{0}}}{\hbar }\right)+\frac{1}{2} \left( 1+e^{\frac{2 x {x_{0}}}{\sigma ^2}}\right)\right]}{\pi  \hbar }
\end{equation}
and
\begin{equation}
\label{wc2}
W_{c_{2}}=N_{2}^{2}\frac{e^{-\frac{2 \sigma ^2 (p + {p_{0}})^2}{\hbar ^2}-\frac{x^2}{2 \sigma ^2}} \left[\frac{1}{2} \left(1+e^{\frac{8 p {p_{0}} \sigma ^2}{\hbar ^2}}\right)+e^{\frac{2 {p_{0}} \sigma ^2 (2 p + {p_{0}})}{\hbar ^2}} \cos \left(\frac{2 {p_{0}} x}{\hbar }\right)\right]}{\pi  \hbar }
\end{equation}
Using Eqs.(\ref{wc1}) and (\ref{wc2}) and further simplifying, we obtain 
\begin{eqnarray} 
\label{wrho}
W_{\rho}&=&(2\pi  \hbar )^{-1}e^{-\frac{2 p^2 \sigma ^2}{\hbar ^2}-\frac{x^2}{2 \sigma ^2}}\big[ N_{1}^{2}  e^{-\frac{x_{0}^2}{2 \sigma ^2}}(e^{-\frac{2 x {x_{0}}}{\sigma ^2}}+e^{\frac{2 x {x_{0}}}{\sigma ^2}})+N_{2}^{2} e^{-\frac{2 p_{0}^2 \sigma ^2}{\hbar ^2}}(e^{-\frac{4 p {p_{0}} \sigma ^2}{\hbar ^2}}+e^{\frac{4 p {p_{0}} \sigma ^2}{\hbar ^2}})\\
\nonumber
&+&2  N_{1}^{2}\cos \left(\frac{2 p x_{0}}{\hbar }\right)+ 2  N_{2}^{2}\cos \left(\frac{2 p_{0} x}{\hbar }\right) \big]
\end{eqnarray}

The Wigner function $W_{\rho}$  is plotted in Fig.1, for a fixed value of $\sigma=0.5 $ cm, $ x_{0}=4.5$ cm, and $p_{0}=10$ gm.cm/s. It can be seen that such a mixed state provides the chess-board type of structure \cite{zurek}. Subsequently, we derive the area of the smallest tile, which gives the information of how sensitively we can perform the measurement of small parameters using the mixed state. In order to obtain the smallest possible tile in the chess-board structure, one has to find the points in the phase-space, where the destructive and constructive interferences occur. This can be obtained from the condition at which $W_{\rho}$ vanishes. From Eq.(\ref{wrho}),  it is clear that $e^{-\frac{2 p^2 \sigma ^2}{\hbar ^2}-\frac{x^2}{2 \sigma ^2}}$ can only vanish at infinity. Hence, the condition when the Wigner function vanishes is given by, 
\begin{equation}
N_{1}^{2} e^{-\frac{x_{0}^2}{2 \sigma ^2}}(e^{-\frac{2 x {x_{0}}}{\sigma ^2}}+e^{\frac{2 x {x_{0}}}{\sigma ^2}})+ N_{2}^{2} e^{-\frac{2 p_{0}^2 \sigma ^2}{\hbar ^2}}(e^{-\frac{4 p {p_{0}} \sigma ^2}{\hbar ^2}}+e^{\frac{4 p {p_{0}} \sigma ^2}{\hbar ^2}})+2 N_{1}^{2} \cos \left(\frac{2 p {x_{0}}}{\hbar }\right)+ 2 N_{2}^{2} \cos \left(\frac{2 {p_{0}} x}{\hbar }\right)=0
\end{equation}
For a fixed $\sigma$,  by taking sufficiently large values of $x_{0}$  and $p_{0}$, the contribution of exponential parts can be neglected and the normalization constants $N_{1}$ and $N_{2}$ become almost equal to  $1/2$. For such choices of $x_{0}$  and $p_{0}$, the values of $x$ and $p$, for which $W_{\rho}$  becomes zero can be obtained from the relation,
\begin{equation}
\cos\left(2 p_{0} x/\hbar \right)+\cos\left(2 x_{0} p/\hbar \right)=0,
\end{equation}
leading to the conditions,  
\begin{equation}
x=\frac{(2n+1)\pi\hbar}{4p_{0}}, \ \ \ p=\frac{(2n+1)\pi\hbar}{4x_{0}},
\end{equation}
where $n=0,1,2...$. All the individual tiles will have the same area with the area of a single tile is;
\begin{equation}
A = x p=\frac{\pi^2\hbar^2}{16x_{0} p_{0}}
\end{equation}
It is seen that by making the values of $x_{0}$ and $p_{0}$ suitably large one can reduce the area to a value as small as one wants. Hence, it is clear that a sub-Planck structure is formed in the phase space for the mixed state. We now investigate the usefulness of this state for metrological purpose for sensitive estimation of small parameters with improved precision.

\section{Sensitivity of the mixed state in estimating small parameters}
In order to find the sensitivity of the mixed state  given by Eq.(\ref{wf1}) for small parameter estimation, we perturb the density matrix $\rho$ by small amounts in both position and momentum co-ordinates. For this, we introduce two phase shifts of $e^{i \delta_1 x}$ and $e^{i \delta_2 p}$ to the states $\psi_{c_{1}}(x)$ and $\psi_{c_{2}}(x)$ respectively, where $\delta_1$ and $\delta_2$ are considered to be very small. For a quantitative measure, how precisely one can estimate the small shifts, the orthogonality between perturbed and unperturbed states are considered \cite{toscano}. Such a procedure, in turn, implies the position and momentum shifts $p+\delta_{1}$ and $x+\delta_{2}$ to the corresponding Wigner function $W_{\rho}$ given by Eq.(\ref{wrho}). The perturbed Wigner function $W_{\rho}^{\prime}$ is given by,

\begin{eqnarray}
W_{\rho}^{\prime}&=&(2\pi  \hbar)^{-1}e^{-\frac{2 (p+\delta_{1})^2 \sigma ^2}{\hbar ^2}-\frac{(x+\delta_{2})^2}{2 \sigma ^2}}[N_{1}^{2} e^{-\frac{x_{0}^2}{2 \sigma ^2}}(e^{-\frac{2 (x+\delta_{2}) {x_{0}}}{\sigma ^2}}+ e^{\frac{2 (x+\delta_{2}) {x_{0}}}{\sigma ^2}})+ N_{2}^{2} \ e^{-\frac{2 p_{0}^2 \sigma ^2}{\hbar ^2}}(e^{-\frac{4 (p+\delta_{1}) {p_{0}} \sigma ^2}{\hbar ^2}}+e^{\frac{4 (p+\delta_{1}) {p_{0}} \sigma ^2}{\hbar ^2}})\\
\nonumber
&+&{2  N_{1}^{2} \cos \left(\frac{2 {x_{0}(p+\delta_{1}) }}{\hbar }\right)+ 2  N_{2}^{2}  \cos \left(\frac{2 {p_{0}} (x+\delta_{2})}{\hbar }\right)}]
\end{eqnarray}
We check the orthogonality between the Wigner functions corresponding to perturbed and unperturbed density matrices by the overlap formula,
\begin{equation}
O=\int_{-\infty}^{\infty}dx\int_{-\infty}^{\infty}dp \ W_{\rho} \ W_{\rho}^{\prime},
\end{equation}
For the density matrices $W_{\rho}$ and $W_{\rho}^{\prime}$ to be orthogonal,  $O=0$. As pointed out earlier, by taking larger value of ${x_{0}}$ and ${p_{0}}$, one can neglect the exponential part. The remaining sinusoidal part of $O$ can be written as, 
\begin{equation}
O=\frac{e^{-\frac{{\delta_{1}}^2 \sigma ^2}{\hbar ^2}-\frac{{\delta_{2}}^2}{4 \sigma ^2}} \left[2 \cos \left(\frac{{2 \delta_{2}} {p_{0}}}{\hbar }\right)+3 \cos \left(\frac{2 \delta_{1} {x_{0}}}{\hbar }\right)+3\right]}{2 \pi  \hbar }
\end{equation}
Orthogonality between $W_{\rho}$ and $W_{\rho}^{\prime}$ requires the values of $\delta_{1}$ and $\delta_{2}$ to be,
\begin{equation}
\delta_{1}=\frac{(2n+1)\pi \hbar}{2 x_{0}} , \ \  \delta_{2}=\frac{(2n+1)\pi \hbar}{4 p_{0}}
\end{equation}
Thus, the smallest possible product of $\delta_1$ and $\delta_2$, we can distinguish by the mixed state is given by, 
\begin{equation}
{\delta_{1}}{\delta_{2}}=\frac{\pi^2 \hbar^2}{8  x_{0} p_{0}},
\end{equation}
This can be achieved for $n=0$. Evidently, if ${x_{0}}$ and ${p_{0}}$ are very large, one can, in principle, estimate very small perturbations.

\section{Preparation of the mixed state and effect of decoherence on it}
The superposition of two coherent states has been experimentally generated for photons\cite{kir} and single trapped ions\cite{monroe}. The mixed state in Eq.(\ref{wf1}) can experimentally be created and manipulated by using a strong nonlinear interaction at the single photon level. In fact, in a recent paper Kirchmair \emph{et al.}\cite{kir} have demonstrated how to experimentally  generate the cat and kitten states using the single-photon Kerr effect. The Hamiltonian describing interaction in Kerr medium is of the following form 
\begin{equation}
{\widehat{H}}_{K}= \frac{\hbar\kappa}{2} (\widehat{a}^{\dagger} \widehat{a})^2  
\end{equation}
where $\widehat{a}$ and $\widehat{a}^{\dagger}$ are annihilation and creation operators respectively and $\kappa$ is the nonlinear constant provides the strength of the interaction. The time evolution of a coherent state $|\alpha\rangle$ can be described as an infinite superposition of different photon number state is one mode Kerr state is given by
\begin{equation}
|\psi_{K}(\tau)\rangle=e^{-i {\widehat{H}}_{K} \tau/\hbar}|\alpha\rangle
\end{equation}
where $\tau=-\kappa t$. For suitable choices of $\alpha$ and $\tau$ one can generate superposition of two or more coherent states. For our purpose, we require the incoherent mixture of the following states $(|\alpha\rangle+|-\alpha\rangle)/\sqrt{2}$ and $(|i\alpha\rangle+|-i\alpha\rangle)/\sqrt{2}$. Note that, the position space projection of those will provide the cat states used in our paper given by Eq.(\ref{psic1}) and Eq.(\ref{psic2}) respectively. We now show the effect of decoherence on the cat states.

In order to show the effect of environment induced decoherence, we let the system interact with a heat bath consisting of a set of harmonic oscillators, initially in equilibrium at temperature ${T=(K_{B}\beta)}^{-1}$. At $t=0$ there is no interaction between the system and the heat bath. The initial Wigner function of the system and environment is given by \cite{halliwell},
\begin{eqnarray}
W_{0} (q,p;q_{i},p_{i})=W_{s} (q,p,0)W_{e} (q_{i},p_{i},0),
\end{eqnarray}
where $W_{s}$ and $W_{e}$ are the Wigner functions of the system and the environment respectively at $t=0$. The Wigner function of heat bath at $t=0$ is taken to be Gaussian,
\begin{eqnarray}
W_{e}=\prod_{n}N_{n}exp[-\frac{2}{\omega_{n}\hbar}tanh\left(\frac{\hbar\omega_{n}}{2}\right)H_{n}]
\end{eqnarray}
where $H_{n}$ is the Hamiltonian of n-th oscillator in the bath with,
\begin{eqnarray}
H_{n}=\frac{p_{n}^2}{2 m}+\frac{m\omega_{n}^2 q_{n}^2}{2}.
\end{eqnarray}
and we consider the same mass for every Harmonic oscillator.

The Wigner function $W(q,p,t)$ of the system plus environment at time $t$ can be obtained by solving the exact master equation \cite{ford};
\begin{eqnarray}
\label{master}
\frac{\partial W}{\partial t} = -\frac{1}{m}P\frac{\partial W}{\partial
q}+m\Omega^{2}(t)q\frac{\partial W}{\partial p} 
+2\Gamma(t)\frac{\partial pW}{\partial p}+\hbar m\Gamma
(t)h(t)\frac{\partial^{2}W}{\partial p^{2}}+\hbar\Gamma
(t)f(t)\frac{\partial^{2} W}{\partial q\partial p},
\end{eqnarray}
where $m$ is mass of harmonic oscillator having frequency $\Omega(t)$, $\Gamma(t)$ is coefficient of quantum dissipation and $\Gamma(t)h(t)$ and $\Gamma(t)f(t)$ are coefficients of quantum diffusion. Solution of Eq.(\ref{master}) with time dependent coefficient is difficult to deal with. The Langevin equation for the initial value problem can be used to find  solution of Eq.(\ref{master}) \cite{gwf}. The general solution of exact master equation in terms of Wigner characteristic function is given by 
\begin{eqnarray}
\label{masterwig}
\tilde{W}(Q,P;t) &=&\exp \left[-\frac{\left\langle X^{2}\right\rangle
P^{2}+m\left\langle X\dot{X}+\dot{X}X\right\rangle QP+m^{2}\left\langle
\dot{ X}^{2}\right\rangle Q^{2}}{2\hbar ^{2}}\right]\\
\nonumber
&\times&\tilde{W}(m\dot{G}Q+GP,m^{2}\ddot{G}Q+m\dot{G}P;0)
\end{eqnarray}
where, $X(t)$ is the fluctuation operator and $G(t)$ is the Green function  
\begin{equation}
 G(t_{2}-t_{1})=\frac{1}{i\hbar}[x(t_{1}),x(t_{2})]\theta (t_{2}-t_{1})
\end{equation}
Here $x(t)$ is the time-dependent Heisenberg coordinate operator and $\theta$ is the Heaviside function. At high temperature, there will be Ohmic model coupling to the heat bath. Assuming the case of free particle moving in the absence of external force and taking frictional coefficient equal to $m \gamma$ \cite{ford}, we have
\begin{equation}
G(t)=\frac{1-e^{-\gamma t}}{m\gamma } 
\end{equation}
The values of the quantities $\left\langle X^{2}\right\rangle=\frac{kT}{m\gamma ^{2}}[2\gamma
t-(1-e^{-\gamma t})(3-e^{-\gamma t})]$, $\left\langle X\dot{X}+\dot{X}X\right\rangle=\frac{2kT}{m\gamma }(1-e^{-\gamma t})^{2}$ and 
$\left\langle \dot{X}^{2}\right\rangle=\frac{kT}{m}(1-e^{-2\gamma t})$ given in Eq.(\ref{masterwig})  are calculated in Ref.\cite{ford}

Considering the initial state given by Eq.(\ref{psic1}), which is a superposition of two minimum uncertainty Gaussian wave packets initially at zero temperature, the Wigner characteristic function is given by
\begin{equation}
\label{wig5}
\tilde{W}(Q,P;t)=\frac{1}{\left(e^{-\frac {x_0^2}{2 \sigma ^2}}+1\right)}\left[e^{-\frac{\frac{4 P^2 \sigma ^4}{\hbar ^2}+Q^2}{8 \sigma ^2}} \left(2 \cos \left(\frac{P x_0}{\hbar }\right)+2 e^{-\frac{x_0^2}{2 \sigma ^2}} \cosh \left(\frac{Q x_0 }{2 \sigma ^2}\right)\right)\right]
\end{equation}
For comparing with the general form given by Eq.(\ref{masterwig}), Eq.(\ref{wig5}) is written as,
\begin{eqnarray}
\tilde{W}(Q,P;t) &=&\frac{1}{1+e^{-x_0^{2}/2\sigma ^{2}}}\exp \{-\frac{
A_{11}^{(0)}P^{2}+2A_{12}^{(0)}PQ+A_{22}^{(0)}Q^{2}}{2\hbar ^{2}}\} 
\nonumber
\\ &&\times (\cos \frac{(m^{2}\ddot{G}Q+m\dot{G}P)x_0}{\hbar
}+e^{-x_{0}^{2}/2\sigma ^{2}}\cosh \frac{(m\dot{G}Q+GP)x_0}{4\sigma ^{2}}),
\end{eqnarray}
where
\begin{equation}
A_{11}^{(0)} =\left\langle X^{2}\right\rangle +\sigma
^{2}m^{2}\dot{G}^{2}+\frac{\hbar ^{2}G^{2}}{4\sigma ^{2}},
\end{equation}
\begin{equation}
A_{12}^{(0)} =m\frac{\left\langle
X\dot{X}+\dot{X}X\right\rangle }{2} +\sigma
^{2}m^{3}\dot{G}\ddot{G}+\frac{\hbar ^{2}mG\dot{G}}{4\sigma ^{2}}, 
\end{equation}
and
\begin{equation}
 A_{22}^{(0)} =m^{2}\left\langle \dot{X}^{2}\right\rangle
+\sigma ^{2}m^{4}
\ddot{G}^{2}+\frac{\hbar ^{2}m^{2}\dot{G}^{2}}{4\sigma ^{2}}. 
\end{equation}
Here the superscript $(0)$ indicates that the particle is at zero temperature before coupling to the heat bath at temperature $T$. In order to calculate $W(q,p;t)$, we use inverse Fourier transform of the form $W(q,p;t)=\frac{1}{(2\pi \hbar )^{2}}\int dQ\int
dPe^{i(qP+pQ)/\hbar }\tilde{W }(Q,P;t)$. Using Eq.(6) we then obtain,
\begin{eqnarray}
\label{wigf1}
 W(q,p;t) &=&\frac{1}{2(1+e^{-(x_{0})^{2}/2\sigma ^{2}})}W^{(0)}(q- m\dot{G}x_{0}, p- m^{2}\ddot{G}x_{0})  \nonumber \\
&& +W^{(0)}(q+ m \dot{G}x_{0}, p+ m^{2}\ddot{G}x_{0})  \nonumber \\
&&+2 e^{-A^{(0)}(t)}W^{(0)}(q,p)\cos \Phi ^{(0)}(q,p:t)
\end{eqnarray}
First two terms of Eq.(\ref{wigf1}) are for the Gaussian wave packets, initially centered at $x_{0}$ and propagating independently and the third term is due to the interference, which is maximum at origin. The quantity $e^{-A^{(0)}(t)}$ is the
measure of interference in phase space \cite{ford}, where  
\begin{eqnarray} 
\label{a0}
A^{(0)}(t) &=&\frac{(A_{11}^{(0)}-\frac{\hbar
^{2}G^{2}}{4\sigma ^{2}} )(A_{22}^{(0)}-\frac{\hbar
^{2}m^{2}\dot{G}^{2}}{4\sigma ^{2}} )-(A_{12}^{(0)}-\frac{\hbar
^{2}m G\dot{G}}{4\sigma ^{2}})^{2}}{
A_{11}^{(0)}A_{22}^{(0)}-A_{12}^{(0)2}}\frac{x_{0}^{2}}{2\sigma ^{2}}
\end{eqnarray}
For $\gamma t\ll 1$, $A^{(0)}$ varies linearly with $t$, $ A^{(0)}(t)\cong \frac{d^{2}}{\lambda
_{th}^{2}}\gamma t$ with $\lambda _{{th}}=\frac{\hbar }{\sqrt{mkT}}$.
Assuming shift $x_{0}$ to be very large as compared to de Broglie wave length,  decoherence time is given by
\begin{equation}
\tau^{1}_{d}=\frac{{\hbar}^2}{4 m \gamma K_{B} T {x_{0}}^2}
\end{equation}
Similarly, for the cat state given by Eq.(\ref{psic2}), the decoherence time can be calculated as
\begin{equation}  
\tau^{2}_{d}=\frac{{\hbar}^4}{16 m \gamma K_{B} T {p_{0}}^2 {\sigma}^4}
\end{equation}
As has been noted earlier, the parameter estimation can be more sensitive for larger values of $x_{0}$ and $p_{0}$. However, we note here that the large value of  $x_{0}$ and $p_{0}$ the decoherence time should be very very small.
\section{summary and conclusions}
The sub-Planck phase-space structures are at the root for the Heisenberg-limited sensitivity, accessible through some specific quantum states like the compass state. We demonstrate that the sub-Planck structure occurs in an incoherent mixture of two cat states. The minimal number of terms in the density matrix are two, which sustain the interference effect in phase space. For the parameter estimation purposes, mixed state performs equally well as compared to the pure compass state. However, the central advantage is that the mixed state is less prone to decoherence than the pure compass state and it is easier to prepare. In particular, in compass state six coherence terms are required to maintain coherence instead of two terms for the case of mixed state. Since the preparation of such mixed state is much easier than the compass state such a mixed state is more useful than the compass state.

\end{document}